\documentclass[twocolumn,aps,prr,longbibliography,superscriptaddress]{revtex4-2}
\usepackage{amsmath}
\usepackage{graphicx}
\usepackage{hyperref}
\usepackage[normalem]{ulem}
\usepackage{bbold}
\hypersetup{colorlinks=true, linkcolor=blue, citecolor=blue, urlcolor=blue}

\begin{document}
\title{Tree-like process tensor contraction for 
automated compression of environments}
%efficient non-Markovian open quantum systems simulations}
\author{Moritz Cygorek}
\affiliation{Condensed Matter Theory, Department of Physics, TU Dortmund, 44221 Dortmund, Germany}
\affiliation{SUPA, Institute of Photonics and Quantum Sciences, Heriot-Watt University, Edinburgh EH14 4AS, United Kingdom}
\author{Brendon W. Lovett}
\affiliation{SUPA, School of Physics and Astronomy, University of St Andrews, St Andrews KY16 9SS, United Kingdom}
\author{Jonathan Keeling}
\affiliation{SUPA, School of Physics and Astronomy, University of St Andrews, St Andrews KY16 9SS, United Kingdom}
\author{Erik M. Gauger}
\affiliation{SUPA, Institute of Photonics and Quantum Sciences, Heriot-Watt University, Edinburgh EH14 4AS, United Kingdom}

\begin{abstract}
The algorithm ``automated compression of environments'' (ACE) [Nat. Phys. 18, 662 (2022)] provides a versatile way of simulating an extremely broad class of open quantum systems. This is achieved by encapsulating the influence of the environment, which is determined by the interaction Hamiltonian(s) and initial states, into compact process tensor matrix product operator (PT-MPO) representations.
The generality of the ACE method comes at high numerical cost. Here, we demonstrate that orders-of-magnitude improvement of ACE is possible by changing the order of PT-MPO contraction from a sequential to a tree-like scheme. The problem of combining two partial PT-MPOs with large inner bonds is solved by a preselection approach. 
The drawbacks of the preselection approach are that the MPO compression is suboptimal and that it is more prone to error accumulation than sequential combination and compression. We therefore also identify strategies to mitigate these disadvantages by fine-tuning compression parameters. This results in a scheme that is similar in compression efficiency and accuracy to the original ACE algorithm, yet is significantly faster.
Our numerical experiments reach similar conclusions for bosonic and fermionic test cases, suggesting that our findings are characteristic of the combination of PT-MPOs more generally.
\end{abstract}
\maketitle

\section{Introduction}
Understanding how quantum systems are affected by their 
environments is a challenging but important task, in particular when the system-environment coupling is not weak and the Born-Markov approximation becomes unreliable~\cite{RevModPhys_deVega}. 
For example, strong interactions between system and environment can shake up environment excitations, which dress and screen the quantum system by polaron formation~\cite{Review_Nazir} or can stabilize localized spins in fermionic impurity models~\cite{Anderson61}.
The richness of phenomena in open quantum systems is further illustrated in the spin-boson model, which features phase transitions between localized and delocalized states as well as regimes with algebraic versus exponential decay~\cite{RevModPhys_Leggett,BreuerPetruccione}, depending on the spectrum of the bosonic bath.
These insights play a crucial role in practical applications, e.g., for understanding phonon effects in semiconductor quantum dots with their typical~\cite{Krummheuer} super-Ohmic spectral density with exponential cut-off.
This coupling explains the appearance of non-Markovian dynamics~\cite{Reiter_distinctive_characteristics2019,Carmele_NonMarkovian2019,
Mogilevtsev2009,Trushechkin2024,PI_QRT}, dynamical decoupling~\cite{Vagov_nonmonotonic,Denning_decoupling,Kaldewey2017,Hall_reappearance}, 
phonon-assisted state preparation~\cite{adiabatic_undressing,PI_singlephoton,Gustin2020}, and
incomplete decay of coherences resulting in long-lived cooperative emission
effects~\cite{CoopWiercinski,CoopSciAdv}.
Even in cases where the decay of excitations or coherences can be described by effective
Markovian master equations~\cite{Jacobs_Lindblad}, the corresponding decay rates and 
Lindblad operators can be challenging to obtain. Because they generally depend 
on system eigenstates and thus on how the system is driven~\cite{Review_Nazir}, they
are particularly difficult to extract in situations with ultrafast time-dependent
driving~\cite{PI_singlephoton, pulsed_spectra} or when system states are nearly
degenerate~\cite{Jacobs_Lindblad,CygorekCoop}.

To resolve such challenging questions in strongly coupled open quantum systems, 
it is highly desirable to have powerful numerical tools to simulate 
complex composite open quantum systems, possibly coupled to several 
different environments, numerically exactly; that is, to approximate exact
results with high precision controlled by convergence parameters without 
imposing additional approximations such as perturbation theory or the
Born-Markov approximation. 

PT-MPOs~\cite{PT_PRA,JP} provide exactly this: They describe the influence of the environment, 
equivalent to the Feynman-Vernon influence functional~\cite{FeynmanVernon}, 
in a compact matrix product operator representation.
Once obtained, they can be used to efficiently simulate the dynamics of the
open quantum system for a given time-dependent system 
Hamiltonian, e.g., to find optimal driving protocols~\cite{Fux_PRL}.
Moreover, multiple PT-MPOs can be stacked together to describe quantum systems
coupled to multiple environments while still remaining 
numerically exact~\cite{ACE,twobath,CoopWiercinski}. 

By now, several algorithms to calculate PT-MPOs have been developed. 
Most are aimed at solving either a generalized 
spin-boson model~\cite{JP,DnC,Strunz_infinite} or quantum impurities coupled to
fermionic baths~\cite{PT_Abanin,PT_Reichman}. In both cases 
Gaussian path integrals can be employed to derive
compact expressions for the nodes of the tensor network to be contracted~\cite{TEMPO}. 
PT-MPOs can also be constructed for more general environments using the 
algorithm \textit{automated compression of environments} (ACE)~\cite{ACE}, which 
starts from a different tensor network:
In this case, one need only assume that the environment is composed of independent modes, which could be, 
e.g., bosons, fermions, spins, or indeed any general non-Gaussian environment.  
Moreover, these modes could be 
time-dependently driven and themselves subject to Markovian loss or dephasing.
For each mode, a PT-MPO is constructed from the respective Hamiltonian.
In ACE, the individual PT-MPOs are then sequentially combined and compressed, 
yielding a PT-MPO encapsulating the joint influence of all modes.
This algorithm is sufficiently abstract that 
it can be implemented by a computer code that is independent of the specific 
physical model at hand. Such a code can serve as a general-purpose solver
to be utilized as a ``black box'' by the user. A C++ implementation is available and described in Ref.~\cite{ACEcode}.

The price to pay for the generality of ACE is the relatively high numerical
cost in comparison to specialized PT-MPO methods for Gaussian environments.
This relative cost is even more significant given recent methodological developments~\cite{DnC,Strunz_infinite} that further reduce the cost for Gaussian environments.
Here we show that ideas used in one of these developments~\cite{DnC} can also lead to a significant speed up for ACE.
Specifically, we will make use of a scheme introduced there to combine and compress MPOs with moderate to large inner bond dimensions by `pre\-selecting' degrees of freedom based on singular value decomposition (SVD) of the individual MPOs.
This ability to combine MPOs with larger bond dimensions enables the contraction of the corresponding tensor network in a different order. 
In particular, the preselection technique facilitates the partial contraction of blocks of the tensor network and the subsequent combination of those blocks at a later step in the algorithm~\cite{DnC}.

In this article, we demonstrate that ACE simulations can be
significantly sped up by using a different contraction order of the tensor network. 
%Concretely, this does not mean a change of the sequence in which individual modes are incorporated into the PT-MPO, for which no consistent speed-up is identified in numerical experiments, as discussed in Appendix~\ref{app:seq}.
Concretely, we suggest to contract PT-MPOs pairwise, yielding 
a tree-like contraction scheme. (A simple reordering of the linear sequence in which modes are contracted provides little improvement, see Appendix~\ref{app:seq}.)
As in Ref.~\cite{DnC}, this scheme is enabled by preselecting 
degrees of freedom when combining and compressing PT-MPOs.
We show several numerical examples for which this tree-like contraction scheme 
significantly outperforms the previous sequential contraction ACE algorithm~\cite{ACE}. 
However, we also observe that  preselection results in suboptimal 
PT-MPO contraction, i.e.~it produces PT-MPOs with larger bond dimensions 
than the sequential ACE algorithm. Furthermore, it accumulates 
larger numerical errors. We show that both of these issues 
can be addressed by fine-tuning convergence parameters, in particular, 
adjusting the number of compression sweeps per MPO combination and 
by choosing a scale-dependent compression threshold.

The article is structured as follows: In Section~\ref{sec:theory} we lay out the theory. We start with a brief summary of the ACE tensor network in Section~\ref{sec:ACE_tensor_network} before we discuss different schemes to contract this network in Section~\ref{seq:schemes}. This includes a brief revision of the sequential contraction scheme of Ref.~\cite{ACE} and of the preselection approach for combining MPOs with medium to large inner bonds from Ref.~\cite{DnC}, followed by the introduction of our tree-like contraction scheme as well as the description of strategies for fine-tuning the MPO compression. 
Numerical results are presented in Section~\ref{sec:results}, where we discuss the model systems in Section~\ref{sec:model}, the computation times required for simulations using different approaches in Section~\ref{sec:times}, the corresponding numerical errors in Section~\ref{sec:err}, and the PT-MPO compression efficacy in Section~\ref{sec:compr_eff}.
Our findings are summarized in the discussion Section~\ref{sec:discussion}.
Further, there are three appendices:
In Appendix~\ref{app:seq}, we briefly investigate the impact of rearranging modes in a different sequence, while in Appendix~\ref{app:fermionic} we provide an elementary derivation of how to obtain fermionic PT-MPOs using ACE. Finally, convergence is discussed in Appendix~\ref{app:norm} using an alternative measure.

\section{Theory\label{sec:theory}}
\subsection{ACE tensor network\label{sec:ACE_tensor_network}}
We first summarize the tensor network that is the basis of ACE~\cite{ACE}.
The dynamics of a quantum system of interest (S) coupled to an environment (E)
is determined by the Liouville-von Neumann equation for the total density matrix
\begin{align}
\frac{\partial}{\partial t}\rho=& \mathcal{L}\rho =
\big(\mathcal{L}_S + \mathcal{L}_E\big)\rho, 
\label{eq:vNeq}
\end{align}
where we split the Liouvillian into a part acting only on the system
degrees of freedom $\mathcal{L}_S$, and an environment part $\mathcal{L}_E$,
which also includes the interaction with the system. These Liouvillians 
are given by the respective system and environment Hamiltonians via
$\mathcal{L}_{S,E} \rho = -\frac i\hbar [H_{S,E}, \rho]$. 
Such a description can be straightforwardly extended to include additional Markovian loss and decoherence channels via Lindbladians as well
as time-dependent Hamiltonians.
It is convenient to work within a superoperator formalism~\cite{inner_bonds}, 
where the density matrix $\rho$ is viewed as a vector and the Liouvillians become
matrices. Then, Eq.~\eqref{eq:vNeq} is formally solved by the matrix exponential
$\rho(t)=e^{\mathcal{L} (t-t_0)} \rho(t_0)$.

The ACE algorithm is obtained by the following steps:
(i)~We introduce a time grid $t_l=t_0 + l\Delta t$ with time steps $\Delta t$
and write $e^{\mathcal{L} t_n}=\prod_{l=1}^n e^{\mathcal{L} \Delta t}$.
(ii)~A Trotter decomposition is employed to split the exponential into 
system and environment parts $e^{(\mathcal{L}_S+\mathcal{L}_E)\Delta t}
=e^{\mathcal{L}_E \Delta t} e^{\mathcal{L}_S \Delta t} + \mathcal{O}(\Delta t^2)$.
(iii)~We assume that at time $t_0$ system and environment are uncorrelated
$\rho(t_0)=\bar{\rho}(t_0)\otimes \rho^E(t_0)$, and trace out the
environment at time $t_n$. This yields
\begin{align}
\label{eq:formaldensmat}
\bar{\rho}(t_n) = &\textrm{Tr}_E\bigg\{ \Big(\prod_{l=1}^{n} 
e^{\mathcal{L}_E \Delta t}e^{\mathcal{L}_S \Delta t}\Big)
\big(\bar{\rho}(t_0) \otimes \rho^E(t_0)\big)\bigg\},
\end{align}
which is cast into the form
\begin{align}
\label{eq:finaldensmat}
\bar{\rho}_{\alpha_n}=&
\sum_{\substack{\alpha_{n-1},\dots,\alpha_0\\\alpha'_n,\dots,\alpha'_1 \\
d_{n-1},\dots,d_1}}
\prod_{l=1}^n \bigg(\mathcal{Q}^{(\alpha_{l},\alpha'_{l})}_{d_l d_{l-1}}
\mathcal{M}^{\alpha'_l \alpha_{l-1}}\bigg) \bar{\rho}_{\alpha_0},
\end{align}
where Liouville space indices $\alpha_l=(\nu_l, \mu_l)$ 
collect left and right (Hilbert space) indices on the 
reduced system density matrix at time $t_l$, i.e.
$\bar{\rho}_{\alpha_l}=\bar{\rho}_{\nu_l,\mu_l}(t_l)$. Furthermore, 
$\mathcal{M}^{\alpha_l,\alpha_{l-1}}=
\big(e^{\mathcal{L}_S\Delta t}\big)_{\alpha_l, \alpha_{l-1}}$ is the free
system propagator while the set of matrices 
$\Big\{ \mathcal{Q}^{(\alpha_l,\alpha'_l)}_{d_l,d_{l-1}}\Big\}$ 
constitute the PT-MPO with inner bonds $d_l$, which carry information about
the state of the environment~\cite{inner_bonds}. 
\begin{figure*}
\includegraphics[width=0.99\textwidth]{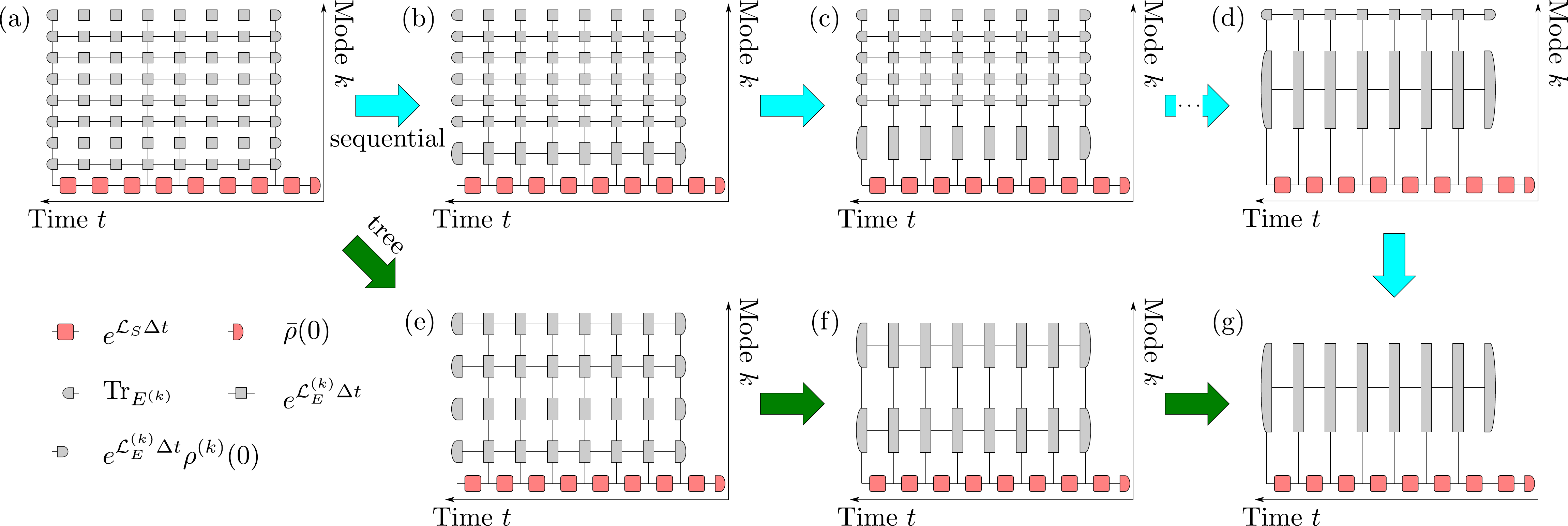}
\caption{\label{fig:sketch}Contraction of the ACE tensor network (a) using
the sequential [(a),(b),(c),(d),(g)] and tree-like [(a),(e),(f),(g)] 
schemes. 
After combining two rows, the inner bonds are compressed by sweeps with
truncated singular value decompositions.
In the sequential scheme, PT-MPOs for individual environment modes
are combined one by one, and so contributions with small inner bond dimensions are added to a PT-MPO with large inner bond dimension.
In the tree-like scheme, neighboring PT-MPOs are combined, so most combination
steps involve the combination of PT-MPOs with small to moderate inner bond
dimension. 
%Only the last steps involve the combination of two PT-MPOs with
%large inner bonds.
Note that panel (a) represents Eq.~\eqref{eq:formaldensmat} with 
$e^{\mathcal{L}_E\Delta t}$ further decomposed as in Eq.~\eqref{eq:envTrotter}.
Panel (g) represents 
Eq.~\eqref{eq:finaldensmat} with the contracted gray blocks corresponding
to PT-MPO matrices $\mathcal{Q}^{(\alpha_l,\alpha'_l)}_{d_l,d_{l-1}}$.
}
\end{figure*}

Comparing Eq.~\eqref{eq:finaldensmat} with Eq.~\eqref{eq:formaldensmat}
one can identify
\begin{align}
\label{eq:Qdef}
&\mathcal{Q}^{(\alpha_l,\alpha'_l)}_{d_l,d_{l-1}} \nonumber\\&=
\begin{cases} 
\big(e^{\mathcal{L}_E \Delta t} (\mathbb{1}\otimes \rho^E(t_0))\big)_{
(\alpha_l,d_l),(\alpha'_l,d_{l-1})}, & l=1,
\\
\big(e^{\mathcal{L}_E \Delta t}\big)_{
(\alpha_l,d_l),(\alpha'_l,d_{l-1})}, &1<l<n
\\
\big(\textrm{Tr}_E\, e^{\mathcal{L}_E \Delta t}\big)_{
(\alpha_l,d_l),(\alpha'_l,d_{l-1})}, & l=n,
\end{cases}
\end{align}
with fixed inner bonds at the boundaries $d_n=d_0=0$ (where we assume indexing from zero), while 
the remaining indices $d_l$ run over a complete basis of the environment 
Liouville space.

Equation~\eqref{eq:finaldensmat}, which is illustrated in Fig.~\ref{fig:sketch}(g),
constitutes a series of matrix multiplications, which propagate the
total density matrix from one time step to the next.
However, direct application of Eq.~\eqref{eq:finaldensmat} is typically 
infeasible because the environment Liouville space is generally extremely 
large. 
This issue can be addressed by tensor network techniques: The set of matrices 
$\big\{\mathcal{Q}^{(\alpha_l,\alpha'_l)}_{d_l,d_{l-1}}\big\}$ 
[gray comb in Fig.~\ref{fig:sketch}(g)] 
has the form of an MPO with inner bonds $d_l$ and 
outer bonds $(\alpha_l,\alpha'_l)$. For such a structure, compression techniques
are available, where the inner bond dimensions are reduced 
while not affecting the action of the total MPO on the outer bonds.
This can be done by sweeping across the MPO from one end to the other
while performing SVDs and eliminating degrees of freedom corresponding to
small singular values below a given threshold~\cite{ACE}.

Because rank-reducing operations typically scale with the third power 
$\mathcal{O}(\chi^3)$ of the inner bond dimension $\chi$, it is 
crucial to start with PT-MPOs with small dimensions and to keep the
bond dimensions manageable at all times.
The sequential ACE algorithm does this by introducing one environment mode at a time.  This implies that this method applies only to environments that are
composed of independent modes, where the environment Liouvillian
$\mathcal{L}^E=\sum_{k=1}^{N_E}\mathcal{L}_E^{(k)}$ can be decomposed into
$N_E$ individual terms $\mathcal{L}_E^{(k)}$. These are assumed to only
affect the system of interest as well as the $k$-th environment mode but they
do not modify other degrees of freedom~\footnote{A modification of the ACE method can also in principle be applied to environments where each mode couples to ``nearest neighbor'' modes, such as arise from chain mapping techniques~\cite{Chin_noise-assisted,T-TEDOPA}.}. Similarly, the initial environment
density matrix is assumed to factorize $\rho^E(t_0)=
\rho^{(1)}(t_0)\otimes \rho^{(2)}(t_0)\otimes \dots\otimes \rho^{(N_E)}(t_0)$.
Then, after employing a further Trotter decomposition of the environment propagator
\begin{align}
\label{eq:envTrotter}
e^{\mathcal{L}_E\Delta t}=&
e^{\mathcal{L}_E^{(1)}\Delta t} e^{\mathcal{L}_E^{(2)}\Delta t} 
\dots e^{\mathcal{L}_E^{(N_E)}\Delta t} +\mathcal{O}(\Delta t^2),
\end{align}
Eq.~\eqref{eq:finaldensmat} becomes the two-dimensional tensor network
depicted in Fig.~\ref{fig:sketch}(a), where each row corresponds to a
PT-MPO for an individual environment mode $k$. If the Hilbert-space dimension
of each individual mode is small enough, the elements of these PT-MPOs can
be directly constructed from Eq.~\eqref{eq:Qdef}. It remains to combine
all the individual PT-MPOs [the rows in Fig.~\ref{fig:sketch}(a)] 
and to compress the inner bonds, such that one eventually obtains the final
PT-MPO in Fig.~\ref{fig:sketch}(g), which then accounts for the influences 
of all environment modes.

\subsection{PT-MPO contraction schemes~\label{seq:schemes}}
Having derived the ACE tensor network, we now discuss schemes
to contract the corresponding PT-MPOs. Generally, combining two PT-MPOs
(two rows in Fig.~\ref{fig:sketch}), whose matrices we denote by
$\mathcal{Q}^{(\alpha_l, \alpha''_l)}_{e_l, e_{l-1}}$ and
$\mathcal{P}^{(\alpha''_l, \alpha'_l)}_{f_l, f_{l-1}}$, respectively, 
yields a new PT-MPO with matrices
\begin{align}
\label{eq:combination}
\tilde{\mathcal{Q}}^{(\alpha_l,\alpha'_l)}_{d_l,d_{l-1}}=&
\sum_{\substack{\alpha''_l\\e_l,e_{l-1}\\f_l,f_{l-1}}}
\delta_{d_l,(e_l,f_l)}\mathcal{Q}^{(\alpha_l, \alpha''_l)}_{e_l, e_{l-1}}
\mathcal{P}^{(\alpha''_l, \alpha'_l)}_{f_l, f_{l-1}}
\delta_{d_{l-1},(e_{l-1},f_{l-1})}.
\end{align}
Here, the Kronecker deltas describe the formation of inner bonds of the combined
PT-MPO as the product of inner bonds of the individual PT-MPOs. 

To counteract the growth of the inner bonds, one compresses the resulting PT-MPO
by sweeping along the MPO chain while performing SVDs. For example, sweeping in the forward
direction (starting from $l=1$), we perform an SVD of the matrix
\begin{align}
A_{d_l, ((\alpha_l,\alpha'),d_{l-1})} =& 
\tilde{\mathcal{Q}}^{(\alpha_l,\alpha'_l)}_{d_l,d_{l-1}} 
\longrightarrow \sum_k U_{d_l, k} \sigma_{k} 
V^\dagger_{k,((\alpha_l,\alpha'),d_{l-1})},
\end{align}
where $U$ and $V$ are matrices with orthonormal columns and $\sigma_k$ are
the non-negative singular values in descending order. Compression is achieved
by keeping only singular values $\sigma_k \ge \epsilon \sigma_0$ 
above a certain threshold $\epsilon$ with respect to the
largest singular value $\sigma_0$.  The PT-MPO matrices at steps $l$ and $l+1$ 
are updated as
\begin{subequations}
\begin{align}
\tilde{\mathcal{Q}}^{(\alpha_l,\alpha'_l)}_{k,d_{l-1}} 
\longleftarrow & V^\dagger_{k,((\alpha_l,\alpha'),d_{l-1})}, \\
\tilde{\mathcal{Q}}^{(\alpha_{l+1},\alpha'_{l+1})}_{d_{l+1},k} 
\longleftarrow & \sum_{d_l} 
\tilde{\mathcal{Q}}^{(\alpha_{l+1},\alpha'_{l+1})}_{d_{l+1},d_l}
\sum_k U_{d_l, k} \sigma_{k}.
\end{align}
\end{subequations}
The same procedure is repeated at the next step (increasing $l$). 
After a forward sweep, one performs an analogous sweep in the backward 
direction.

It should be noted that such a compression is only proven to be optimal if the MPO is of 
canonical form with orthogonality center at the site to be compressed~\cite{MPS_Schollwoeck}.
This is generally not the case if the MPO is the result of combining two MPOs 
via Eq.~\eqref{eq:combination}. In principle, a sweep along the chain in 
one direction without truncation brings the MPO in such a form. However, in the
spirit of the zip-up algorithm of Ref.~\cite{MPS_StoudenmireWhite}, it is
common practice in PT-MPO algorithms~\cite{JP} to perform truncation in
every sweep because this reduces computation times while often providing
close-to-optimal compression.

The central question we address in this article is the performance of
different schemes to contract the PT-MPO 
for the network in Fig.~\ref{fig:sketch}(a). 
We consider the following variants:

\subsubsection{Original (sequential) ACE}
In the original ACE algorithm in Ref.~\cite{ACE}, the tensor network is
contracted by including the influence of one mode at a time sequentially 
into a growing PT-MPO. This sequence is depicted in Fig.~\ref{fig:sketch} 
along panels (a), (b), (c), (d), and (g).
After including each additional environment mode, a single forward and backward sweep 
is performed to compress the inner bonds with a fixed compression threshold
$\epsilon$.

The sequential order was chosen in Ref.~\cite{ACE} because of the scaling
of SVDs with the third power of the matrix dimension. Together with the
fact that combining two PT-MPO matrices as in Eq.~\eqref{eq:combination}
results in multiplication of the respective inner bond dimensions, brute-force
calculations of SVDs become intractable if both matrices have large inner dimension 
(in practice the product of inner bonds each $\gtrsim 20$ prove challenging).

\subsubsection{Preselected singular values}
The problem of combining two PT-MPOs, both with moderate to large inner bond 
dimensions also arose in the divide-and-conquer scheme for PT-MPOs with Gaussian (bosonic) reservoirs Ref.~\cite{DnC}. 
In that work we addressed this issue by `preselecting' relevant parts of matrices before
performing the most demanding SVDs. 

This preselection is done by first performing independent forward sweeps of the two PT-MPOs
%matrices $\mathcal{Q}^{(\alpha_l, \alpha''_l)}_{e_l, e_{l-1}}$
%and $\mathcal{P}^{(\alpha''_l, \alpha'_l)}_{f_l, f_{l-1}}$ 
to be combined.
The corresponding singular values $\sigma^{(1)}_{e_l}$ and $\sigma^{(2)}_{f_l}$
are stored and used as indicators for the numerical relevance of the
row $e_l$ and $f_l$, respectively. The PT-MPOs are then combined during the
subsequent backward sweep, where we keep
only those indices $d_l$ for which the products of singular
values $\sigma^{(1)}_{e_l}\sigma^{(2)}_{f_l} \ge \epsilon
\sigma^{(1)}_{0}\sigma^{(2)}_{0}$ exceed a threshold. This threshold is defined
relative to the 
product of the maximal individual singular values (and likewise for $d_{l-1}$).
This preselection step dramatically reduces the workload for the backward sweep SVD that follows
the combination. 

The combination of two moderately-sized PT-MPOs via the preselection approach 
enables new algorithms. However, this approach removes information from
the MPO before compression, and so it can lead to a less effective compression
than found for the zip-up approach alone. Thus, numerical tests are needed to investigate 
the practicality of preselecting degrees of freedom, which may depend on the
specific example considered~\cite{DnC}. 

In this article we explore one variant of tensor network contraction in which we follow 
the same sequential contraction order as in the original ACE algorithm, yet
combine pairs of PT-MPOs using preselection based on prior individual SVDs.  This comparison lets us separate the effects of the MPO-MPO contraction from the effects of contraction order.

\subsubsection{Tree-like contraction}
The original sequential ACE contraction always only adds a small influence---%
resulting from one individual mode---at a time. However, this implies that
it requires operations involving one PT-MPO with large inner dimensions for most
of the combination and compression steps. For example, in the penultimate step
in Fig.~\ref{fig:sketch}(d), the inclusion of the last mode, which by itself
should have a small effect, requires operations on a PT-MPO with inner bonds
of the size comparable to the full final PT-MPO.  
This can in principle be avoided by other contraction orders.

The preselection method now allows us to explore a different contraction order,
where we first combine neighboring modes, assuming a sequence where these have similar influences (e.g., when modes are listed in order of increasing mode energies; see Appendix~\ref{app:seq} for discussion of alternate orderings). 
The resulting PT-MPOs are then further combined, leading to a tree-like
overall contraction order depicted in Fig.~\ref{fig:sketch} as the sequence
along panels (a), (e), (f), and (g).  While both, sequential ACE 
and the tree-like contraction, require $N_E-1$ combinations and compressions,
the distribution of inner bond dimensions is drastically different.
In particular, in the tree-like scheme, the first $N_E/2-1$ contractions 
involve combinations with the smallest possible inner bond dimensions.
If neighboring modes have similar physical influences, 
the combined PT-MPOs should be compressible to small dimensions, which also
makes the next layer of $N_E/4-1$ combinations and compressions efficient.
Only the last few combination and compression steps involve PT-MPOs
with large inner bond dimensions.
This is the key argument why the tree-like contraction scheme can be expected
to outperform contraction in the sequential order.

Finally, it is noteworthy that the tree-like scheme is in principle parallelizable 
except for the last step. As our goal is to compare the algorithms themselves,
we do not make use of this in our present study.

\subsubsection{Fine-tuning of MPO compression}
The preselection method can in general result in suboptimal 
MPO compression~\cite{DnC}. 
We therefore explore two additional strategies to mitigate this.
The first strategy is based on the following observation: 
Even when MPOs are locally optimally compressed by sweeps with 
truncated SVDs starting from MPOs in canonical form, 
truncations further into the chain depend on earlier truncations closer
to the starting edge. Hence, better compression may be achieved by sweeping 
multiple times $n_\textrm{SW}>1$ after combining two PT-MPOs.
In many-body quantum physics, this is often done using iterative, e.g.,
variational, methods~\cite{MPS_Schollwoeck}, which are faster than 
repeated SVD sweeps~\footnote{Nevertheless, most applications in quantum
many-body physics still require an initial SVD line sweep to provide a 
starting point for an iterative scheme~\cite{MPS_Schollwoeck}.}.
Motivated by this, we check
for potential improvements by sweeping multiple times 
$n_\textrm{SW}>1$ with truncating SVDs after combining two PT-MPOs.

The second strategy is to reduce the accumulation of rounding errors 
by choosing scale-dependent compression, where different
compression thresholds are used for different layers of the tree-like scheme. 
Concretely, we start with a compression threshold $\epsilon_\textrm{min}$ for
the first layer and exponentially increase the threshold for each successive layer, 
such that the last compression is done with the maximal threshold 
$\epsilon_\textrm{max}$. This corresponds to equidistant sampling of $\ln\epsilon$ in the range  $[\ln\epsilon_\textrm{min},\ln\epsilon_\textrm{max}]$.
This way, excess degrees of freedom resulting from suboptimal compression
at earlier stages of the algorithm have a chance to be eliminated in later 
steps, while computation time is increased mainly at the first few steps,
which are the fastest steps in the tree-like scheme.
We parameterize the scale-dependent compression by the threshold range factor 
$r=\epsilon_\textrm{max}/\epsilon_\textrm{min}$ and identify 
$\epsilon_\textrm{max}=\epsilon$ with the nominal compression threshold 
$\epsilon$, e.g., for comparison with sequential ACE simulations.

\subsection{Reducing Trotter errors}
For clarity, we have so far described the ACE tensor network 
assuming a first-order (asymmetric) Trotter split between system and environment 
as well as between different environment modes. However, the Trotter error
can be reduced~\footnote{If a fixed final time $t$ is discretized on a grid with 
$n$ steps of size $\Delta t=t/n$, the total Trotter error scales as
$\mathcal{O}(1/n)$ for the first-order and $\mathcal{O}(1/n^2)$ for the
second-order Trotter formula, respectively, because Trotter errors occur at 
each of the $n$ time steps.}
from $\mathcal{O}(\Delta t^2)$ to 
$\mathcal{O}(\Delta t^3)$ using a second-order (symmetric) Trotter decompostion of the form
\begin{align}
e^{(A+B)\Delta t} =&e^{A \Delta t/2} e^{B\Delta t} e^{A \Delta t/2}
+\mathcal{O}(\Delta t^3).
\end{align}
The system-environment split can be straightforwardly brought into symmetric
form without changing the PT-MPO by calculating system propagators over
half times steps $e^{\mathcal{L}_S \Delta t/2}$.

The Trotter decomposition between different environment modes in Eq.~\eqref{eq:envTrotter}
can also be easily implemented using the symmetric split in the sequential contraction scheme, where
the PT-MPO is expanded by including the environment propagator for a single 
environment mode. There, one can replace Eq.~\eqref{eq:combination} by a 
symmetric version where half-step propagators $e^{\mathcal{L}_E^{(k)}\Delta t/2}$ 
are multiplied from the left as well as from the right to the current PT-MPO
(see Ref.~\cite{ACE} for details).

In contrast, in the tree-like scheme, implementing the symmetric Trotter 
splitting is more challenging because the time steps of PT-MPOs in 
intermediate layers of the algorithm are fixed. We solve this issue
by changing the order of combination~\cite{inner_bonds}: At odd time steps
(starting from $l=1$), we combine PT-MPO matrices in the same order 
as in Eq.~\eqref{eq:combination}, whereas at even time steps, we change
the order (with respect to the outer indices $\alpha$) to
\begin{align}
\label{eq:combination2}
\tilde{\mathcal{Q}}^{(\alpha_l,\alpha'_l)}_{d_l,d_{l-1}}=&
\sum_{\substack{\alpha''_l\\e_l,e_{l-1}\\f_l,f_{l-1}}}\delta_{d_l,(e_l,f_l)}
\mathcal{P}^{(\alpha_l, \alpha''_l)}_{f_l, f_{l-1}}
\mathcal{Q}^{(\alpha''_l, \alpha'_l)}_{e_l, e_{l-1}}
\delta_{d_{l-1},(e_{l-1},f_{l-1})}.
\end{align}
Then, the propagation over two steps effectively amounts to a symmetric 
Trotter splitting, where the lower-order Trotter error introduced by an odd
time step is canceled by the corresponding term in the subsequent even step.

\section{Results\label{sec:results}}
\subsection{Model systems\label{sec:model}}
\begin{figure*}
\includegraphics[width=0.99\textwidth]{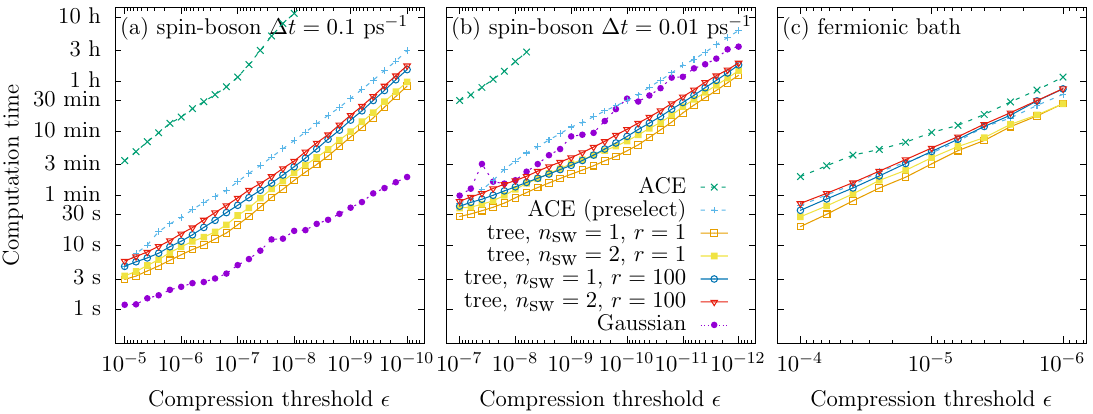}
\caption{\label{fig:times}Computation times 
of PT-MPO contraction for a spin-boson model with larger (a) and 
smaller (b) time steps, and for a fermionic environment (c) 
using different methods: The original sequential ACE algorithm~\cite{ACE} (green), the sequential algorithm using
the preselection approach for PT-MPO combination and compression (light blue),
and the tree-like scheme using $n_{SW}=1$ or 2 SVD sweeps per combination and
a scale-dependent compression threshold with range factor $r=1$ and $r=100$, 
respectively. For the spin-boson model, we compare with the algorithm of
J{\o}rgensen and Pollock in Ref.~\cite{JP} which assumes Gaussian 
environment correlations (purple).
Note that the scale for both computation times 
and compression thresholds $\epsilon$ is logarithmic. 
Note also that thresholds $\epsilon$ decrease from left
to right on the axis, so that accuracy improves from left to right.
}
\end{figure*}

We test the performance of the different variants for contracting the ACE tensor network on two examples.

First, we consider a two-level system coupled to a bath of harmonic
oscillators, i.e. the paradigmatic spin-boson model. For this model,
very efficient alternative PT-MPO construction algorithms are 
available~\cite{JP,DnC,Strunz_infinite}, which can be used as a 
reference. These make use of the Gaussian statistics of 
bath correlations, which facilitates the PT-MPO construction using a
different tensor network derived by path integration~\cite{TEMPO}, which in turn is based on previous path integration approaches~\cite{QUAPI1,QUAPI2}.
Concretely, the environment Hamiltonian is given by $H_E=\sum_k H_E^{(k)}$ 
with
\begin{align}
H_E^{(k)}=&\hbar\omega_k b^\dagger_k b_k +
\hbar g_k (b^\dagger_k + b_k) |e\rangle\langle e|
+\hbar\frac{g_k^2}{\omega_k}|e\rangle\langle e|,
\end{align}
where $b^\dagger_k$ and $b_k$ are bosonic creation and annihilation operators,
$\hbar\omega_k$ is the energy of the $k$-th environment mode, $g_k$ are the
coupling constants and $|e\rangle\langle e|$ is the projection onto the
excited state of the two-level spin system. The last term is added for convenience
to cancel to reorganization energy or polaron shift.

We choose environment parameters $g_k, \omega_k$ to reproduce a particular form of the spectral density 
$J(\omega)= \sum_k |g_k|^2 \delta(\omega-\omega_k)$. Specifically, we choose parameters such that $J(\omega)$ takes the form of a well established model~\cite{Krummheuer} for a 
quantum dot coupled to longitudinal acoustic phonons 
\begin{align}
J(\omega)
%=& \frac{\omega^3}{4\pi^2 \rho\hbar c_s^5}
%\bigg(D_e e^{-\omega^2 a_e^2/(4c_s^2)} - D_h e^{-\omega^2 a_h^2/(4c_s^2)}
%\bigg)^2.
=& \omega^3 \bigg(c_e e^{-\omega^2/\omega_e^2} - c_h e^{-\omega^2/\omega_h^2}
\bigg)^2.
\end{align}
For a GaAs quantum dot with radius 4 nm relevant parameters are $c_e=0.1271$ ps$^{-1}$, $c_h=-0.0635$ ps$^{-1}$,
$\omega_e=2.555$ ps$^{-1}$, and $\omega_h=2.938$ ps$^{-1}$.

For the ACE tensor network, we discretize the frequency range  
$\omega=[0,\omega_\textrm{max}]$ uniformly, with $\omega_\textrm{max}=7\textrm{ ps}^{-1}$ and using $N_E=64$ modes. Thus, the mode frequencies are then $\omega_k=(k-\frac 12)\omega_\textrm{max}/N_E$ and the coupling strengths are $g_k=\sqrt{J(\omega_k)\omega_\textrm{max}/N_E}$ for $k=1,\dots, N_E$. 
The initial state of each mode is given by a thermal distribution
with temperature $T=4$ K. We check convergence with respect to the
Hilbert-space dimension per mode, and find $M=4$ 
(up to three bosonic excitations) is sufficient. A total propagation time of $t_e=20$ ps 
is discretized with steps of size $\Delta t=0.1$ ps or $\Delta t=0.01$~ps.
The system Hamiltonian $H_S$ does not enter the PT-MPO calculation algorithm.

The second example we consider is a fermionic resonant level model. This
is relevant, e.g., for impurity problems~\cite{PT_Reichman}. The environment
Hamiltonians are given by
\begin{align}
H_E^{(k)}=&\hbar\omega_k c^\dagger_k c_k +
\hbar g_k (c^\dagger_k c_0 + c_0^\dagger c_k ),
\label{eq:fermionic}
\end{align}
where again $\hbar\omega_k$ and $g_k$ are mode energies and coupling constants,
respectively. The operators  $c^\dagger_k$ and $c_k$ are fermionic 
creation and annihilation operators, where the mode $k=0$ is the system mode while modes $k>0$ constitute the environment. 

Here, we model the hopping of electrons from metallic leads to a central site. If the band width of the environment density of states is finite, the dynamics is in general non-Markovian. For concreteness, we assume that the environment density of states is centered around the system mode energy, which we shift to zero---such a shift has no effects for particle-number-conserving models. We thus discretize the (dimensionless) energy range $\omega\in[-32, 32]$ uniformly with $N_E=128$ modes. The coupling constants $g_k=\sqrt{64/(2\pi N_E)}$ are chosen such that the hopping rate
in the Markov limit is $\Gamma=1$. All environment modes are initially in the occupied 
state, which describes a situation where the central site is 
energetically far below the bath's Fermi level.
A time grid spanning the time $t_e=5$ in steps of size $\Delta t=0.1$ is used for
the calculations.

It should be noted that the derivation of ACE assumes that operators corresponding to different environment modes commute. By contrast, fermionic operators anticommute. In Appendix~\ref{app:fermionic}, we derive how one can nevertheless obtain PT-MPOs properly accounting for the fermionic anticommutation relations using ACE. This only requires a slight modification of the input to the algorithm. %, and is otherwise similar to the calculation of a PT-MPO for an analogous open quantum system with fermionic operators in Eq.~\eqref{eq:fermionic} replaced by spin-1/2 ladder operators. 
This strategy can be applied irrespective of whether the contraction order is sequential or tree-like.

\subsection{Computation times\label{sec:times}}
In Fig.~\ref{fig:times} we show the computation times needed
to calculate PT-MPOs for the spin-boson model with larger ($\Delta t=0.1$~ps, panel~a) and 
smaller ($\Delta t=0.01$~ps, panel~b) time steps, as well as for the fermionic model (panel~c).
The different variants of contracting the ACE tensor network described in the 
theory section are used. The computation times
are extracted as the total elapsed time on a conventional laptop computer 
with AMD Ryzen 7 5825U processor.
\begin{figure}
\includegraphics[width=0.99\linewidth]{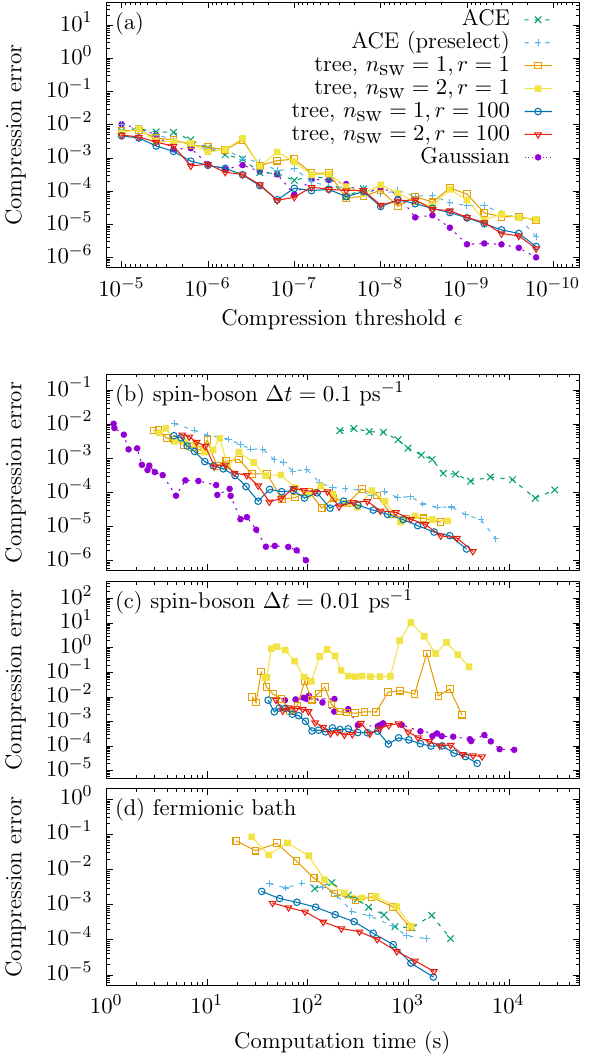}
\caption{\label{fig:err}(a): Compression error (see main text) as a function of the compression threshold for the same simulations as in 
Fig.~\ref{fig:times}(a).
(b), (c), and (d): Compression error versus computation time for the same simulations as in Fig.~\ref{fig:times}(a), (b), and (c), respectively.}
\end{figure}

In both models, we obtain similar trends, which remain robust over
the range of thresholds shown in Fig.~\ref{fig:times}:
sequential ACE simulations using the original algorithm of Ref.~\cite{ACE} require
by far the longest computation times and become impractical when high 
accuracies (small thresholds $\epsilon$) are needed. 
Using the sequential algorithm but with the preselection approach for combining 
rows of the tensor network already leads to massive savings in computation times.
In the spin-boson example in Fig.~\ref{fig:times}(a) the computation time 
at $\epsilon=10^{-7}$ is reduced
by a factor of about $40$, whereas we find a speed-up by a factor 2 
in the fermionic model in Fig.~\ref{fig:times}(c) at $\epsilon=10^{-5}$. 
This is due to the fact that the larger Liouville space per mode in
the spin-boson model ($M^2=16$ in the spin-boson model versus $M^2=4$ for fermions) 
has more potential for reduction by a prior forward sweep before combination.

Switching to the tree-like scheme, we find a further reduction of 
computation time with the fastest simulations being those without any fine-tuning
of convergence parameters, i.e., sweeping only once ($n_{SW}=1$) per combination 
and using a fixed compression threshold $r=1$. 
For the spin-boson example, the advantage over the original ACE algorithm is roughly two orders of magnitude.
Compared to the method by J{\o}rgensen and Pollock~\cite{JP}---which is restricted to Gaussian environments, hence labeled as `Gaussian' in Fig.~\ref{fig:times}---the more general tree-like ACE scheme requires longer computation times  
for those simulations with larger time steps [Fig.~\ref{fig:times}(a)]. 
However, when time steps are smaller [Fig.~\ref{fig:times}(b)], sequential ACE is faster than the specialized J{\o}rgensen and Pollock algorithm.
This is because the latter scales quadratically with the number of time steps,
while the tree-like scheme is linear in $n$~\footnote{Note however that there exist quasilinear~\cite{DnC} 
and linear-scaling~\cite{Strunz_infinite} PT-MPO methods for Gaussian environments}.

\subsection{Numerical error\label{sec:err}}
The numerical error due to MPO compression is analyzed in Fig.~\ref{fig:err}(a). 
To do this, we simulate the dynamics of a two-level system driven with a 
system Hamiltonian $H_S=(1$~ps$^{-1})\frac{\hbar}2 \sigma_x$ and coupled to 
the bosonic environment described by the PT-MPOs calculated in 
Fig.~\ref{fig:times}(a). We define the compression error as the maximal 
(over all time steps) difference between 
occupations $n_e(t_j)=\langle \big(|e\rangle\langle e|\big)\rangle_{t_j}$ 
from simulations with compression threshold $\epsilon$ and those from
simulations with the smallest threshold in each data series ($\epsilon=10^{-8}$ for the sequential ACE without preselection and $\epsilon=10^{-10}$ for all other methods).

In Fig.~\ref{fig:err}(a), it can be seen that, for the same compression threshold $\epsilon$, the tree-like contraction scheme results in
larger errors than the Gaussian algorithm unless a scale-dependent 
compression threshold ($r=100$) is used. 
The fact that a similarly large error is found in the sequential ACE scheme with 
preselection indicates that this error is due to the preselection rather than the contraction order. 

Especially relevant for practical applications is the trade-off between compression error and computation time for each of the methods considered here, which is depicted in Fig.~\ref{fig:err}(b), (c), and (d) for the same simulations as in Fig.~\ref{fig:times}(a), (b), and (c), respectively. It is clearly seen that, first of all, the tree-like contraction outperforms the sequential versions of ACE with and without preselection, and second, using a scale-dependent compression threshold leads to improved accuracy for the same computation time in the tree-like scheme. Moreover, the Gaussian algorithm outperforms all ACE variants when the time steps are large [Fig.~\ref{fig:err}(b)], but the tree-like ACE scheme with scale-dependent compression threshold becomes more efficient than the Gaussian algorithm for small time steps [Fig.~\ref{fig:err}(c)]. 

It should be noted that very small time steps $\Delta t$ can introduce numerical artifacts. This is because in environment propagators $e^{\mathcal{L}_E \Delta t}\approx \mathbb{1} +\mathcal{L}_E\Delta t + \dots$ the information from $\mathcal{L}_E$ is scaled by $\Delta t$, which results in numerical values much smaller than the unity elements of the identity matrix and thus to numerical cancellation effects. Moreover, for the same overall propagation length, calculations with small times steps require many more individual combination and compression steps, and so numerical errors can accumulate. This can lead to unstable, nonconverging behavior of the tree-like scheme when error accumulation is not mitigated by a scale-dependent compression threshold [see Fig.~\ref{fig:err}(c)]. Additionally, due to the long computation times, we cannot extract reliable estimates for the compression error using sequential ACE without preselection. With preselection, the sequential ACE scheme suffers even worse numerical instabilities than the tree-like scheme, leading to unphysical results with occupations greater than unity for almost all thresholds $\epsilon$ considered [not shown in Fig.~\ref{fig:err}(c)]. We have checked that also the sequential ACE with preselection produces reliable results when an scale-dependent compression threshold $r=100$ is used. 
The accuracy per computation time for the fermionic bath shown in Fig.~\ref{fig:err}(d) behaves qualitative similar to that of the spin-boson model and also benefits from the use of a scale-dependent compression threshold. 
%Interestingly, at high accuracies (corresponding to large bond dimensions) simulations using multiple sweeps $n_\textrm{SW}>1$ are found to be beneficial. 

In Appendix~\ref{app:norm}, we analyze the same situation as in Fig.~\ref{fig:err}(a) and (b) but determining the convergence error via a distance measure of MPOs. This measure is independent of the choice of system Hamiltonian and yields smoother convergence curves, but numerical cancellation prevents its use very close to convergence. 

\subsection{Compression efficacy\label{sec:compr_eff}}

\begin{figure}
\includegraphics[width=0.99\linewidth]{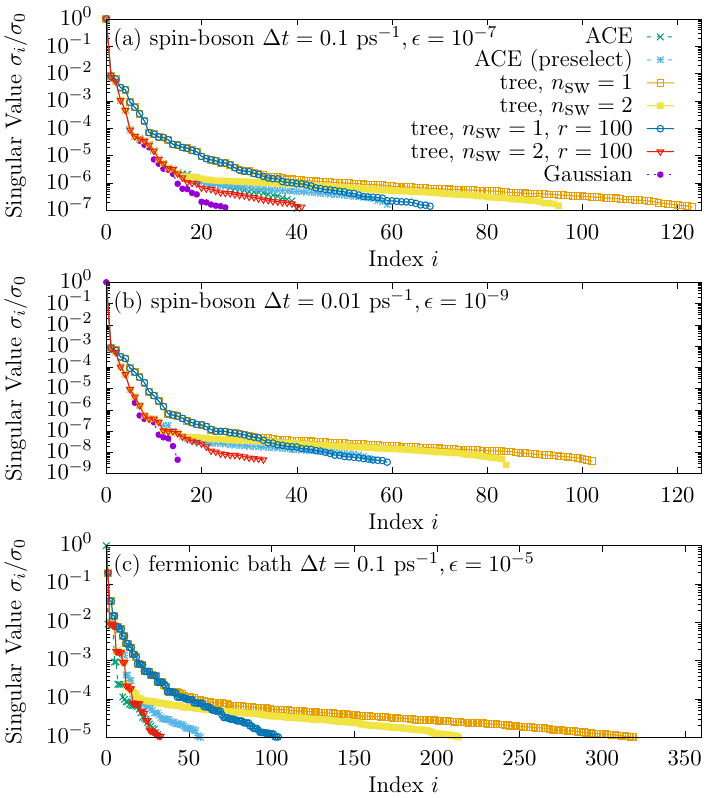}
\caption{\label{fig:sv} Normalized singular values at the center of the final PT-MPO
for the spin-boson model with large (a) and small (b) times steps, and (c) for the fermionic environment, 
corresponding to calculations in Fig.~\ref{fig:times}(a), (b), and (c), respectively.
}
\end{figure}

The final performance measure we discuss is how effective the reduction of 
environment degrees of freedom is, which can be measured by the
inner bond dimension $\chi$ of the PT-MPO for a given compression threshold
$\epsilon$. To study this we analyze the distribution of singular values of the matrix 
$\mathcal{Q}^{(\alpha_l,\alpha'_l)}_{d_l,d_{l-1}}$ 
in the middle ($l=n/2$) of the final PT-MPOs that result from different contraction 
algorithms. This is shown in Fig.~\ref{fig:sv} for  the
spin-boson model with large time steps $\Delta t=0.1$ ps at compression threshold $\epsilon=10^{-7}$,
for the spin-boson model with small time steps $\Delta t=0.01$ ps at threshold $\epsilon=10^{-9}$,
and for the fermionic environment at threshold $\epsilon=10^{-5}$.

As a reference, in Fig.~\ref{fig:sv}(a) we also show singular values obtained by the Gaussian
algorithm of Ref.~\cite{JP}, which typically results in
close-to-optimal compression~\cite{Strunz_infinite}. Here, we obtain 
26 singular values above the threshold. In contrast, the original sequential ACE algorithm 
results in 41 singular values, where the largest ones agree with those
from the Gaussian algorithm.
Using ACE with preselection yields 60 singular values and
the tree-like contraction algorithm without fine-tuning produces 124. 
Singular values obtained from tree-like contraction deviate visibly 
from the close-to-optimal ones already from the fourth value (index $i=3$),
and the tree-like algorithm produces many singular values lying in a
long tail within about one order of magnitude of the truncation threshold.

As can be seen in Fig.~\ref{fig:sv}(a),
the larger singular values can be brought close to the values of the Gaussian
method by sweeping multiple times $n_\textrm{SW}>1$ per PT-MPO combination. 
Furthermore, by comparing singular values from the tree-like scheme without and with
scale-dependent thresholds ($r=100$), we find that the latter strongly
suppresses the long tail of small singular values. Along with the results 
on the numerical error, this suggests that the small singular values are
spurious and a result of error accumulation.
However, the scale-dependent threshold does not affect the large singular values,
which still do not match the close-to-optimal values.
This shows that both strategies for fine-tuning compression parameters address 
different aspects of suboptimal PT-MPO compression in the tree-like contraction scheme.
Indeed, bringing both strategies together---i.e. sweeping twice and
dynamically increasing the threshold---yields a similar number of singular
values as the original ACE algorithm.
As shown in Fig.~\ref{fig:times}(a), the computation times are however still reduced by one to two orders of magnitude compared to sequential ACE.
Similar results are found for both the spin-boson model with small time steps in Fig.~\ref{fig:sv}(b)
and the fermionic environment in Fig.~\ref{fig:sv}(c).

\begin{figure}
\includegraphics[width=0.99\linewidth]{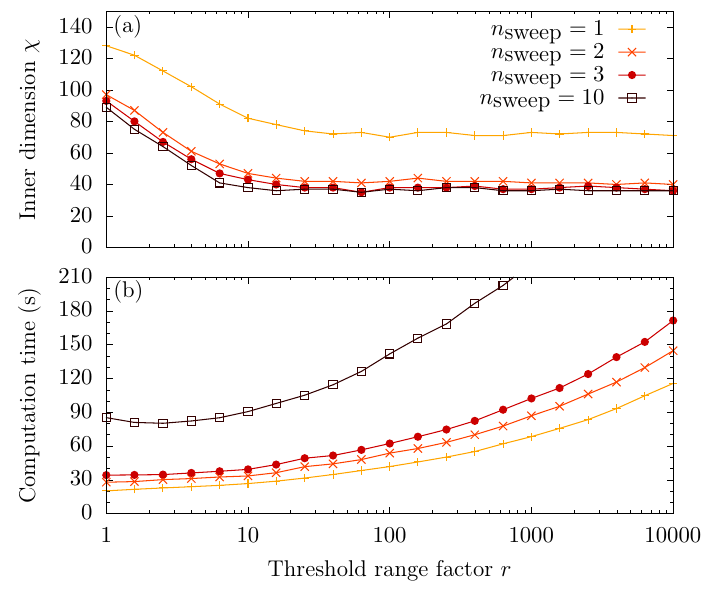}
\caption{\label{fig:vs_tr}Inner dimension $\chi$ (a) and computation time (b)
for different numbers of MPO compression sweeps $n_\textrm{SW}$ per combination 
as a function of the threshold range factor $r=\epsilon_\textrm{max}/\epsilon_\textrm{min}$
for the scale-dependent compression threshold. 
Simulations were done for the same parameters as in Fig.~\ref{fig:times}(a) with 
$\epsilon_\textrm{min}=\epsilon=10^{-7}$.
}
\end{figure}

Finally, Fig.~\ref{fig:vs_tr} shows in more detail how inner bond dimensions
and computation times depend on the fine-tuning parameters, i.e. 
the number of sweeps $n_\textrm{SW}$ and the threshold range factor $r$, on
the example of the spin-boson model from Fig.~\ref{fig:times}(a).
We find that sweeping twice per PT-MPO combination significantly reduces bond 
dimensions, but increasing the number of sweeps $n_\textrm{SW}$ 
further has only marginal effect on the compression.
Similarly, increasing the factor $r$ controlling the scale-dependent 
compression threshold initially reduces the bond dimensions, but these then saturate when $r\approx 10$.
The computation time increases with $n_\textrm{SW}$ and with $r$, yet
the increase with $r$ is relatively slow. 
Thus, for a given environment, experimenting with the large values of $r$ to
find a good trade-off between accuracy, compression, and computation time
is manageable.

\section{Discussion\label{sec:discussion}}
We have introduced and analyzed a novel process tensor contraction scheme, with which the ACE algorithm~\cite{ACE} for the simulation of open quantum systems can be considerably improved. The new, tree-like contraction order requires a way to combine and compress
PT-MPOs with large inner dimensions, for which we use the preselection approach
described in Ref.~\cite{DnC}. 
Our scheme achieves about two orders of magnitude reduction in 
computation time compared to the original sequential ACE algorithm. 
The reason for this speed-up is the typical distribution of
inner bond dimensions in different steps of the algorithm: most steps 
in the sequential scheme involve contracting an object with a small bond dimension with a PT-MPO with a large inner dimension; in contrast the tree-like scheme combines influences of similar dimensions, which are
on average smaller than those of the final PT-MPO.
However, the compression---as defined by the the reduction
of bond dimension and numerical error---is suboptimal for the tree-like algorithm, unless the compression parameters are fine-tuned.
We have explored multiple line sweeps with SVDs after combining two PT-MPOs as well as
the use of a scale-dependent compression threshold;  we indeed find that these strategies give a significant 
reduction of inner bond dimensions and of numerical errors. 

Even though the exact value of optimal fine-tuning parameters may depend on the concrete
physical problem, the general trends found in our examples suggest the following 
rules for choosing them: 
As calculations without fine-tuning ($n_\textrm{SW}=1, r=1$) are fastest, one
should first perform such calculations, from which one can also extract and analyze
the distribution of singular values in the final PT-MPO. If these show a long tail 
of small singular values near the truncation threshold, it is advisable 
to increase $r$ in steps corresponding to orders of magnitude (i.e., $r=10, 100, 1000, \dots$) 
until no significant reduction in the number of singular values in the final PT-MPO is found.
Sweeping multiple times $n_{\textrm{SW}}>1$ per PT-MPO combination is often not necessary if one is only 
interested in numerical accuracy and computation time. However, sweeping twice $n_{\textrm{SW}}=2$
can help if close-to-optimal MPO compression is sought.  
Close-to-optimal compression is important in several cases; these include: when the simulation complexity 
encoded in the inner bond dimension $\chi$ is of interest, when the PT-MPO is reused for many
calculations, or when quantum systems with multiple environments described by multiple PT-MPOs
will be simulated. In any case, sweeping more than twice is typically not profitable.

Summarizing, we find that the tree-like contraction scheme clearly outperforms the ACE algorithm
with sequential contraction over all test cases considered, yet it remains generally applicable. 
Thus, we expect that our new scheme also significantly accelerates investigations of 
complex open quantum systems including those coupled to non-Gaussian environments, 
or those which are manipulated with time-dependent driving and are lossy. 
Moreover, all insights gained here relate to the combination of multiple PT-MPOs in general.
This is also important in different setting, e.g., for combining two PT-MPOs describing
different environments~\cite{CoopWiercinski} and 
their non-additive cross effects~\cite{Nazir_nonadditive,twobath}. 
Furthermore, the algorithms described here can be useful to parallelize PT-MPO schemes, e.g., where different parts 
of the ACE tensor network are contracted on different computing nodes before they are combined to form
the final PT-MPO.

\acknowledgements
M.C.~acknowledges funding by the Return Program of the State of North Rhine-Westphalia. M.C.~and
E.M.G.~acknowledge funding from EPSRC grant no. EP/T01377X/1. 
B.W.L.~and J.K.~acknowledge funding from EPSRC grant no. EP/T014032/1.

\appendix
\section{\label{app:seq}Impact of sequential ordering of modes}
\begin{figure}
\includegraphics[width=0.999\linewidth]{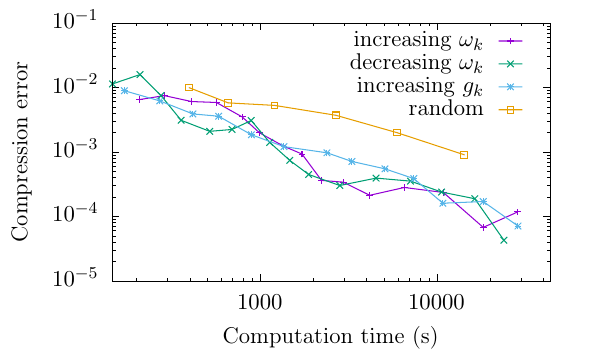}
\caption{\label{fig:order}Compression error vs.~computation time for sequential ACE simulations of the same spin-boson model as in Fig.~\ref{fig:err}(b), where the environment modes are sorted in order of increasing frequency, decreasing frequency, and increasing coupling strength, respectively. For comparison we also show results found from a single random ordering of modes.}
\end{figure}

In the main text, we discuss the improvement of the ACE algorithm by switching from a sequential to a tree-like order of contractions. The question naturally arises whether another aspect of contraction order, namely reordering the individual modes within the sequential scheme, has the potential for improving the efficiency of the ACE algorithm. 

To investigate this question, we perform convergence studies of the sequential ACE algorithm for the spin-boson model discussed in Fig.~\ref{fig:err}(a) and (b), where we sort the environment modes in order of increasing mode frequency $\omega_k$, decreasing mode frequency $\omega_k$, and increasing coupling strength $g_k$. Additionally, we compare to the situation with randomly enumerated modes. The results are shown in Fig.~\ref{fig:order} and indicate no consistent preference for choosing any particular of the ordered sequences. However, the results do indicate that randomly ordering the modes results in much longer computation times for a given compression error. 
This observation can be understood by noting that random ordering inhibits efficient compression of modes with similar influences at intermediate steps of the algorithm.  For definiteness, in the simulations presented in the main text of the article, we choose to label the modes in increasing order in frequency, i.e. $\omega_{k_1}<\omega_{k_2}$ for $k_1<k_2$.

\section{\label{app:fermionic}Fermionic PT-MPOs}
In the main text, we consider the fermionic open quantum system with
total Hamiltonian $H_E=\sum_{k=1}^{N_E}H_E^{(k)}$, where
\begin{align}
H_E^{(k)}=\hbar\omega_k c^\dagger_k c_k +
\hbar g_k (c_k^\dagger c_0 + c^\dagger_0 c_k ).
\end{align}
Here, $c^\dagger_j$ and $c^{}_j$ are fermionic creation and annihilation 
operators, respectively, which obey the canonical anti-commutator 
relations $\{c^{}_i,c^\dagger_j\}=\delta_{ij}$ and  $\{c_i,c_j\}=0$.
We can compare this to the spin-like system with two-level
environment modes described by 
\begin{align}
\tilde{H}_E^{(k)}=\hbar\omega_k \sigma^+_k \sigma^-_k +
\hbar g_k (\sigma_k^+ \sigma^-_0 + \sigma^+_0 \sigma_k^-),
\label{eq:spinmodel}
\end{align}
where $\sigma^+_j$ and $\sigma^-_j$ are two-level (spin-1/2) ladder 
operators. These also satisfy $\{\sigma^-_j,\sigma^+_j\}=1$ 
for a single mode $j$, but spin ladder operators for different modes 
commute $[\sigma^-_i,\sigma^\pm_{j\neq i}]=0$. That is, no sign change occurs when
changing the order of spin ladder operators, while it does for fermions.

While the spin-like system is directly in a form that allows a numerically
exact treatment using ACE, this is not straightforward for a proper fermionic
model, where anticommutation relations have to be enforced.
Here, we show how the fermionic model can be mapped to a spin-like model,
which turns out to be a slight variation of Eq.~\eqref{eq:spinmodel}.
This provides an alternative approach to recent
%Our short and elementary derivation circumvents the technical complexity involved in 
derivations of fermionic tensor network techniques using
Grassmann variables~\cite{PT_Abanin,PT_Reichman,Guo_Grassmann}.

We first employ the Jordan-Wigner transformation to replace 
fermionic operators by spin operators
\begin{subequations}
\begin{align}
c^\dagger_j &= P^{[0,j-1]} \sigma^+_j,\\
c_j &= P^{[0,j-1]} \sigma^-_j,
\end{align}
\end{subequations}
where 
\begin{align}
P^{[j_1,j_2]}=\prod_{j=j_1}^{j_2} (-\sigma^z_j)
\end{align}
will be referred to as the parity operator for modes $j_1$ to $j_2$.
It is straightforward to show that the respective sets of fermionic and spin operators obey the 
corresponding (anti)commutation relations. Applying this
transformation to the fermionic environment Hamiltonian, we obtain
\begin{align}
H_E^{(k)}=\hbar\omega_k \sigma^+_k \sigma^-_k +
\hbar g_k P^{[1,k-1]} (\sigma_k^+ \sigma^-_0 + \sigma^+_0 \sigma_k^-).
\end{align}
In contrast to the spin system defined by $\tilde{H}_E^{(k)}$, here, 
the parity operator $P^{[1,k-1]}$ 
introduces a non-local coupling between different environment modes.
However, progress can be made by realizing that both Hamiltonians are 
connected by
\begin{align}
H_E^{(k)}=\big(P^{[1,k-1]}\big)^{\sigma^+_0\sigma^-_0} 
\tilde{H}_E^{(k)}\big(P^{[1,k-1]}\big)^{\sigma^+_0\sigma^-_0},
\end{align}
which can be seen by noting that 
$\big(P^{[1,k-1]}\big)^{\sigma^+_0\sigma^-_0}$ commutes with $\sigma^{\pm}_k$ 
and $\big(P^{[1,k-1]}\big)^{\sigma^+_0\sigma^-_0} \sigma^{\pm}_0 \big(P^{[1,k-1]}\big)^{\sigma^+_0\sigma^-_0} = P^{[1,k-1]}\sigma^{\pm}_0$.
Because $\Big[\big(P^{[1,k-1]}\big)^{\sigma^+_0\sigma^-_0}\Big]^2=1$, we may see that $\big(P^{[1,k-1]}\big)^{\sigma^+_0\sigma^-_0}$ is a unitary transform.
Hence, we can relate the time evolution operators for both models by
\begin{align}
e^{\pm \frac{i}{\hbar} H_E^{(k)} \Delta t} &=
\big(P^{[1,k-1]}\big)^{\sigma^+_0\sigma^-_0}
e^{\pm \frac{i}{\hbar} \tilde{H}_E^{(k)} \Delta t}
\big(P^{[1,k-1]}\big)^{\sigma^+_0\sigma^-_0}.
\end{align}

While the fermionic and spin Hamiltonians for each mode $k$ are related by a 
unitary transform, this transform is different for each $k$, so the total fermionic and spin Hamiltonians are not unitarily equivalent.
However, we can nonetheless combine the time evolution operators for the different environment
modes as follows.  The standard Trotter decomposition gives
\begin{widetext}
\begin{align}
e^{\pm \frac{i}{\hbar} \sum_{k} H_E^{(k)} \Delta t}
&\approx e^{\pm \frac{i}{\hbar} H_E^{(N_E)} \Delta t}
\dots e^{\pm \frac{i}{\hbar} H_E^{(3)} \Delta t}\,
e^{\pm \frac{i}{\hbar} H_E^{(2)} \Delta t}\,
e^{\pm \frac{i}{\hbar} H_E^{(1)} \Delta t}
\nonumber\\&= 
%\big(P^{[1,N_E-1]}\big)^{\sigma^+_0\sigma^-_0} 
%e^{\pm \frac{i}{\hbar} \tilde{H}_E^{(N_E)} \Delta t}
%\big(P^{[1,N_E-1]}\big)^{\sigma^+_0\sigma^-_0} 
\dots 
\big(P^{[1,2]}\big)^{\sigma^+_0\sigma^-_0} 
e^{\pm \frac{i}{\hbar} \tilde{H}_E^{(3)} \Delta t}
\big(P^{[1,2]}\big)^{\sigma^+_0\sigma^-_0} 
\big(P^{[1,1]}\big)^{\sigma^+_0\sigma^-_0} 
e^{\pm \frac{i}{\hbar} \tilde{H}_E^{(2)} \Delta t}
 \big(P^{[1,1]}\big)^{\sigma^+_0\sigma^-_0} 
e^{\pm \frac{i}{\hbar} \tilde{H}_E^{(1)} \Delta t}
\nonumber\\&= 
\big(P^{[1,N_E]}\big)^{\sigma^+_0\sigma^-_0}
\bigg[ \big(P^{[N_E,N_E]}\big)^{\sigma^+_0\sigma^-_0}
e^{\pm \frac{i}{\hbar} \tilde{H}_E^{(N_E)} \Delta t} \bigg]\dots
\bigg[ \big(P^{[1,1]}\big)^{\sigma^+_0\sigma^-_0}
e^{\pm \frac{i}{\hbar} \tilde{H}_E^{(1)} \Delta t} \bigg].
\label{eq:Trotter_wide}
\end{align}
\end{widetext}
For the step from the second to the last line, we use the fact that
$P^{[1,j+1]}P^{[1,j]}=P^{[j+1,j+1]}$. This results in cancellation of most 
of the non-local parity terms, so that the full time evolution 
almost completely decouples into a product of ``local'' terms, 
each acting only on a single environment mode as well as the system mode. 
Each of these local terms---which are collected into square brackets---have the form of
the time evolution for a single mode spin-like system,
$e^{\pm \frac{i}{\hbar} \tilde{H}_E^{(k)} \Delta t}$ multiplied by 
$\big(P^{[k,k]}\big)^{\sigma^+_0\sigma^-_0}$, the local parity operator.
This parity operator is $-1$ if both environment and system mode are simultaneously occupied 
and $1$ otherwise. 
 
The only remaining non-local term is the parity operator of the full 
environment $\big(P^{[1,N_E]}\big)^{\sigma^+_0\sigma^-_0}$.
This term can however be eliminated using a Trotter decomposition with alternating
order over two time steps. That is, for even time steps $n$ (starting from $n=0$),
we use the Trotter decomposition
\begin{align}
e^{\pm \frac{i}{\hbar} (H_E+H_S) \Delta t}
&\approx e^{\pm \frac{i}{\hbar} H_E^{(N_E)} \Delta t}
\dots e^{\pm \frac{i}{\hbar} H_E^{(1)} \Delta t}\,
e^{\pm \frac{i}{\hbar} H_S \Delta t},
\end{align}
as above, while for odd time steps $n$ we use
\begin{align}
e^{\pm \frac{i}{\hbar} (H_E+H_S) \Delta t}
&\approx  e^{\pm \frac{i}{\hbar} H_S \Delta t}\,
e^{\pm \frac{i}{\hbar} H_E^{(1)} \Delta t}
\dots e^{\pm \frac{i}{\hbar} H_E^{(N_E)} \Delta t}.
\end{align}
Then, the factor $\big(P^{[1,N_E]}\big)^{\sigma^+_0\sigma^-_0}$ appearing on
the left of the decomposition for even $n$ (see Eq.~\eqref{eq:Trotter_wide}) 
exactly cancels with the corresponding factor 
appearing on the right of the subsequent (odd $n$) time step.
A side effect of the alternating order of Trotter decompositions is
that it also cancels the leading-order Trotter error~\cite{inner_bonds}.
The key point of this approach is that the time evolution over two time steps 
now completely decouples into terms involving only single environment modes. 
This makes the fermionic problem amenable to a treatment with the ACE algorithm.

Having established the mapping above, we conclude this appendix by discussing the practical implementation of the resulting fermionic PT-MPOs.
For the construction of the fermionic PT-MPOs, 
it should be noted that the PT-MPO matrices can be understood
as the environment propagator acting on the full environment Liouville space 
compressed to the most relevant subspace of environment excitation, where the
lossy compression matrices and their pseudoinverses are denoted by 
$\mathcal{T}$ and $\mathcal{T}^{-1}$, respectively. 
These matrices are determined implicitly
by the combination and rank-reducing compression~\cite{inner_bonds} in the
sequential or the tree-like ACE algorithm.
For the spin-like system, using the superoperator formalism of Ref.~\cite{inner_bonds}, the PT-MPO take the form
\begin{align}
\label{eq:Qspin}
\tilde{\mathcal{Q}}=&
\mathcal{T} e^{\tilde{\mathcal{L}}_E\Delta t} \mathcal{T}^{-1}=
\mathcal{T}\bigg[ e^{-\frac{i}{\hbar}\tilde{H}_E \Delta t}\otimes 
\big(e^{\frac{i}{\hbar}\tilde{H}_E \Delta t}\big)^T \bigg]\mathcal{T}^{-1}
\nonumber\\=&
\mathcal{T}\bigg[  \prod_{k=1}^{N_E}
e^{-\frac{i}{\hbar}\tilde{H}_E^{(k)} \Delta t}\otimes 
\big(e^{\frac{i}{\hbar}\tilde{H}_E^{(k)} \Delta t}\big)^T \bigg]\mathcal{T}^{-1}
,
\end{align}
where $(\ldots)^T$ indicates transpose.
For the fermionic system, using the Jordan-Wigner transform, the PT-MPO matrices instead take the form:
\begin{align}
{\mathcal{Q}}=&
\mathcal{T}\bigg[  \prod_{k=1}^{N_E}\Big(
\big(P^{[k,k]}\big)^{\sigma^+_0\sigma^-_0}
e^{-\frac{i}{\hbar}\tilde{H}_E^{(k)} \Delta t}\Big)
\nonumber\\&
\otimes \Big(
e^{\frac{i}{\hbar}\tilde{H}_E^{(k)} \Delta t}
\big(P^{[k,k]}\big)^{\sigma^+_0\sigma^-_0}
\Big)^T \bigg]\mathcal{T}^{-1}.
\label{eq:Qmapped}
\end{align}

By comparison, we see that the fermionic PT-MPO can be obtained in the 
same way as the PT-MPO of the corresponding spin system.
One simply has to multiply the main input of the algorithm---the
environment propagators for the individual environment modes 
$e^{\tilde{\mathcal{L}}_E^{(k)}\Delta t}$---by the local parity operator
$\big(P^{[k,k]}\big)^{\sigma^+_0\sigma^-_0}$ in both, forward and backward
propagation direction.

\section{\label{app:norm}Convergence using tensor distance}

\begin{figure}
\includegraphics[width=0.999\linewidth]{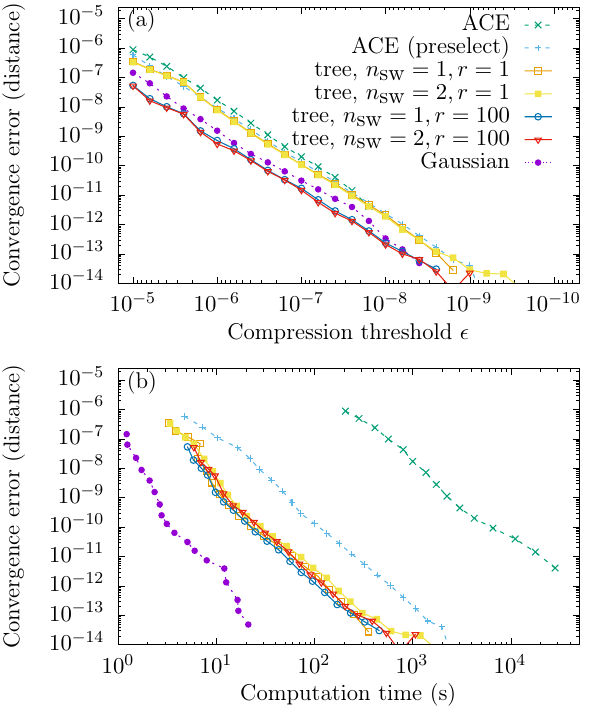}
\caption{\label{fig:norm}Convergence error in terms of the tensor distance as a function of (a) compression threshold $\epsilon$ and (b) computation time. Parameters are chosen as in Fig.~\ref{fig:err}(a) and (b).}
\end{figure}

In the main text, we analyzed the convergence error for a specific system observable and system Hamiltonian. Alternatively, one may be interested in a global measure for convergence that is independent of the choice of observable and system propagator. Given two PT-MPOs $A$ and $B$, we consider the normalized distance
\begin{align}
d(A,B)=\frac{|A-B|^2}{|A||B|}=
\frac{\langle A|A\rangle + \langle B|B\rangle - 2\,\textrm{Re}\{\langle A|B\rangle\}}{\sqrt{\langle A|A\rangle \langle B|B\rangle}},
\end{align}
where, labeling the PT-MPO matrices of $A$ and $B$ by $\mathcal{Q}^{[A]}$ and $\mathcal{Q}^{[B]}$, respectively, the overlap is calculated by contracting over the outer bonds
\begin{align}
\langle A|B\rangle=&
\sum_{\substack{\alpha_{n-1},\dots,\alpha_0\\\alpha'_n,\dots,\alpha'_1 }}
\sum_{\substack{d_{n-1},\dots,d_1\\
\tilde{d}_{n-1},\dots,\tilde{d}_1}}
\prod_{l=1}^n \bigg(\mathcal{Q}^{[A](\alpha_{l},\alpha'_{l})}_{d_l d_{l-1}}
{\mathcal{Q}}^{[B](\alpha_{l},\alpha'_{l})}_{\tilde{d}_l \tilde{d}_{l-1}}\bigg).
\end{align}
Due to the MPO structure of the PTs, the distance can be calculated in $\mathcal{O}\big(n D^4 (\chi_A^2\chi_B + \chi_A \chi^2_B)\big)$ floating point operations, where $\chi_A$ and $\chi_B$ are the inner bonds of PT-MPOs $A$ and $B$, respectively, and $D$ is the system Hilbert space dimension.

In Fig.~\ref{fig:norm}, we perform the analogous convergence analysis as for the spin-boson model in Fig.~\ref{fig:err}(a) and (b) but using the distance between PT-MPOs calculated with nominal thresholds $\epsilon$ and the PT-MPO with the smallest threshold in the series.
The convergence is similar to that found in Fig.~\ref{fig:err}, where the tree-like scheme provides orders-of-magnitude speed-up over sequential ACE, and using a scale-dependent threshold results in improved accuracy for the same nominal compression threshold $\epsilon$. Compared to Fig.~\ref{fig:err}, the analysis using the tensor distance produces much smoother convergence curves. However, we find that at small thresholds there is significant numerical cancellation, which results in an unreliable measure. The values can even becomes negative, here for $\epsilon \gtrsim 10^{-9}$, which is why no data points are shown for these values in Fig.~\ref{fig:norm}.

Moreover, it should be noted that the distance between two PT-MPOs makes no quantitative statement about the accuracy of any physical quantity. We thus suggest that, despite the dependence on a concrete system propagator, analyzing the convergence behavior of a given physical observable as done in the main text is more useful for determining the accuracy of PT-MPOs in practical applications.

%\bibliography{references.bib}

\end{document}